\documentclass{jaa}
\usepackage{natbib}
\usepackage{xcolor}
\usepackage{multicol}
\usepackage{soul}
\usepackage[switch,pagewise]{lineno}
\linenumbers

\usepackage{hyperref}
\hypersetup{
    colorlinks=true,
    linkcolor=blue,
    filecolor=magenta,      
    urlcolor=cyan,
    pdftitle={Overleaf Example},
    citecolor=red,
    }
 
\usepackage{graphicx}

\begin{document}\sloppy
\nolinenumbers
\title{High-Performance Computing for SKA Transient Search:\\ Use of FPGA based Accelerators - a brief review}
\author{
Rubana Aafreen\textsuperscript{1},   
R. Abhishek\textsuperscript{2},
B. Ajithkumar\textsuperscript{3},
Arunkumar M. Vaidyanathan\textsuperscript{1}, 
Indrajit. V. Barve\textsuperscript{4},\\
Sahana Bhatt\textsuperscript{2},
Shashank Bhat\textsuperscript{2},
B. S. Girish\textsuperscript{2},
Yashwant Gupta\textsuperscript{1},
Atul Ghalame\textsuperscript{5},\\
Harshal G. Hayatnagarkar\textsuperscript{1},  
P. A. Kamini\textsuperscript{2},
A. Karastergiou\textsuperscript{6},
S. Madhavi\textsuperscript{2},
M. Mekhala\textsuperscript{3},\\
Preeti Malakar\textsuperscript{8}
Mitch Mickaliger\textsuperscript{5},
V. Mugundhan\textsuperscript{2},
Arun Naidu\textsuperscript{6},
Julian Oppermann\textsuperscript{7}\\
Arul Pandian\textsuperscript{2},
Narendranath Patra\textsuperscript{9}
Lina Preston\textsuperscript{5},
A. Raghunathan\textsuperscript{2}, 
Jayanta Roy\textsuperscript{3*},\\
Shiv Sethi\textsuperscript{2},
Benjamin Shaw\textsuperscript{5},
Krys Sherwin\textsuperscript{7}
Oliver Sinnen\textsuperscript{7}
K. S. Srivani\textsuperscript{2},\\
Benjamin Stappers\textsuperscript{5*},
C.R. Subrahmanya\textsuperscript{2},
Prabu Thiagaraj\textsuperscript{2*},
C. Vinutha\textsuperscript{2},\\
Yogesh G. Wadadekar\textsuperscript{3},
Haomiao Wang\textsuperscript{7}\and
Chris Williams\textsuperscript{6}}

\author{
R. Aafreen\textsuperscript{1},   
R. Abhishek\textsuperscript{2},
B. Ajithkumar\textsuperscript{3},
Arunkumar M. Vaidyanathan\textsuperscript{1},
Indrajit. V. Barve\textsuperscript{4},\\
Sahana Bhattramakki\textsuperscript{2}, 
Shashank Bhat\textsuperscript{2},
B. S. Girish\textsuperscript{2},
Atul Ghalame\textsuperscript{5},
Y. Gupta\textsuperscript{1}, \\
Harshal G. Hayatnagarkar\textsuperscript{1},   
P. A. Kamini\textsuperscript{2},
A. Karastergiou\textsuperscript{6},
L. Levin\textsuperscript{5},
S. Madhavi\textsuperscript{2},\\
M. Mekhala\textsuperscript{3},
M. Mickaliger\textsuperscript{5},
V. Mugundhan\textsuperscript{2},
Arun Naidu\textsuperscript{6},
J. Oppermann\textsuperscript{7},
B. Arul Pandian\textsuperscript{2}, \\
N. Patra\textsuperscript{9},
A. Raghunathan\textsuperscript{2}, 
Jayanta Roy\textsuperscript{3*},
Shiv Sethi\textsuperscript{2},
B. Shaw\textsuperscript{5},
K. Sherwin\textsuperscript{7}
O. Sinnen\textsuperscript{7}\\
S. K. Sinha\textsuperscript{10}
K. S. Srivani\textsuperscript{2},
B. Stappers\textsuperscript{5*},
C. R.Subrahmanya\textsuperscript{2}, 
Thiagaraj Prabu\textsuperscript{2*},
C. Vinutha\textsuperscript{2},
Y. G. Wadadekar\textsuperscript{3},
Haomiao Wang\textsuperscript{7}\and
C. Williams\textsuperscript{6}}

\affilOne{\textsuperscript{1} Engineering for Research, Thoughtworks Technologies, Pune, India.{}}
\affilTwo{\textsuperscript{2} Raman Research Institute, Bengaluru, India.{}}
\affilThree{\textsuperscript{3} NCRA, Pune, India.{}}
\affilFour{\textsuperscript{4} IIA, Bengaluru, India.{}}
\affilFive{\textsuperscript{5} University of Manchester, UK.{}}
\affilSix{\textsuperscript{6} University of Oxford,  UK. {}}
\affilSeven{\textsuperscript{7} University of Auckland, New Zealand.{}}
\affilNine{\textsuperscript{9} IIT Indore, Indore, India.{}}
\affilTen{\textsuperscript{10} IISc, Bengaluru, India.{}}


\twocolumn[{

\maketitle

\corres{{prabu@rri.res.in;  jroy@ncra.tifr.res.in;  Ben.Stappers@manchester.ac.uk; }}

\msinfo{X Jan 2023 revised/accepted}{X YYY ZZZZ}

\begin{abstract}
This paper presents the High-Performance computing efforts with FPGA for the accelerated pulsar/transient search for the SKA.  Case studies are presented from within SKA and pathfinder telescopes highlighting future opportunities. It reviews the scenario that has shifted from offline processing of the radio telescope data to digitizing several hundreds/thousands of antenna outputs over huge bandwidths, forming several 100s of beams, and processing the data in the SKA real-time pulsar search pipelines. A brief account of the different architectures of the accelerators, primarily the new generation Field Programmable Gate Array-based accelerators, showing their critical roles to achieve high-performance computing and in handling the enormous data volume problems of the SKA is presented here.  It also presents the power-performance efficiency of this emerging technology and presents potential future scenarios. 
\end{abstract}

}]




\doinum{12.3456/s78910-011-012-3}
\artcitid{\#\#\#\#}
\volnum{000}
\year{0000}
\pgrange{1--}
\setcounter{page}{1}
\lp{1}

\section{Introduction}
\hfill

Since the World War-II, astronomers have been  probing the sky, also in wave-length windows other than that of the visible light. The radio waves that emanate from very distant celestial objects, when received on earth, are extremely weak, noise-like, and their presence can only be detected by very sensitive receivers associated with radio telescopes through a minute modification of statistical characteristics of noise. Such tiny modifications to receiver's noise (often called “signals” in radio astronomy), in general, require sensitive telescopes, use of large integration (exposure) and many times wide bandwidths for reliable detection. These aspects also make the radio astronomy signal detection very vulnerable to radio frequency interference.

The Square Kilometre Array (SKA) is  the most sensitive Radio Telescope, to be deployed in the most radio-quiet parts of the world: African and Australian deserts. The first phase of the project, SKA1 (we refer to as SKA for brevity in the rest of the paper), will commission 197 dish antenna arrays in South Africa and 1,31,072 element aperture arrays in Western Australia. Together, these arrays will observe over an extensive frequency range from 50 MHz to about 15 GHz. These arrays will produce very high volume imaging and non-imaging data streams and support to explore a variety of radio astronomy problems \citep{ron}.  

\begin{figure*}[!ht]
    \centering
  \includegraphics[height= 10cm,width= 17cm]{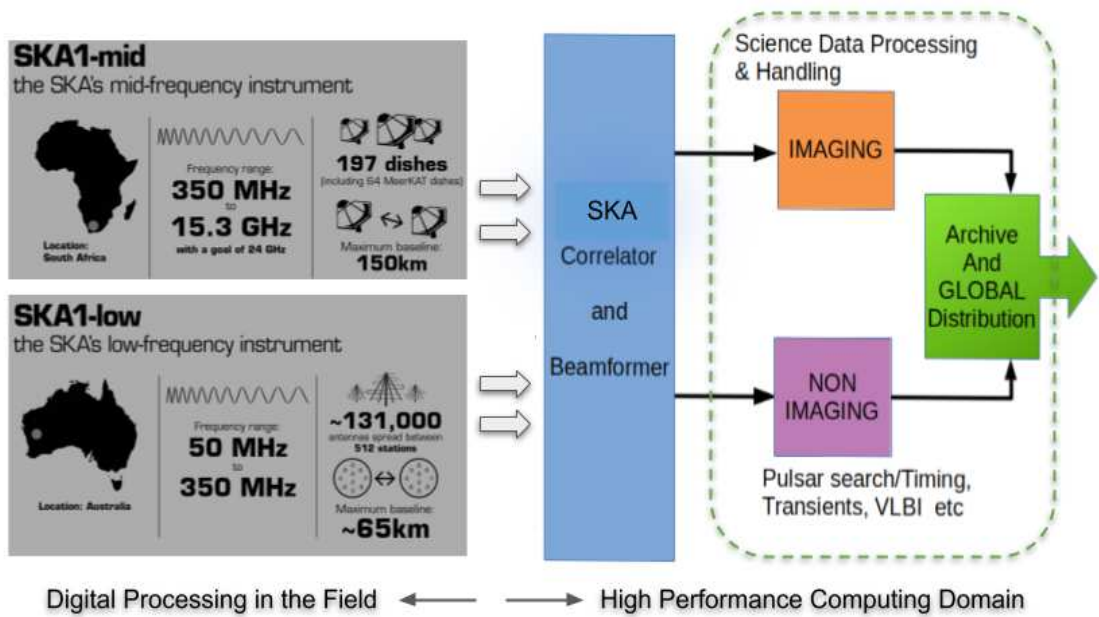}
        \caption[]{ Signal flow overview of SKA}
        \label{fig:ska}
\end{figure*}
The signal flow scheme for the SKA\footnote{\url{www.skatelescope.org}} is presented from a high-performance computing prospective in Fig.~\ref{fig:ska}. Antenna signals are conditioned, digitized and transported to the correlator-beamformers. Outputs from imaging and non-imaging pipelines are distributed/archived using a global network. These pipelines work in soft real-time and use high-performance compute clusters. After this first level of processing,  the end products would need further analysis by the SKA user community in a manner suitable for different science applications. The data volumes are huge as well as the user community will be geographically distributed. Hence the SKA data will be delivered to a set of  SKA Regional Centres (SRCs) located across the globe. The regional science centres will host compute resources to provide access to the SKA data products,  data analysis tools and processing power to the SKA user community.

Typical high-performance computing problems in radio astronomy come from but are not limited to:
\begin{itemize}
    \itemsep0em
    \item Modeling 
    \item Imaging
    \item Multi-beamforming 
    \item Determination of data quality 
    \item Real-time processing of data streams and
    \item Machine learning for: 
    \begin{itemize}
        \itemsep0em
        \item Candidate detection
        \item Interference (RFI)  detection, and
        \item Forecasting applications.
    \end{itemize}    
\end{itemize}

One of the primary projects planned with the SKA is studying and surveying transient objects, such as pulsars. SKA has two kinds of pulsar observation modes: 1) timing the pulsar rotation and 2) searching for fast transients - pulsars and fast radio burst events. While pulsar timing and search are complex and the telescopes have traditionally used offline processing, the SKA will deal with these observations in real-time by using dedicated high-performance processing pipelines. 

\begin{figure*}[!ht]
    \centering
    \includegraphics[height= 8cm,width= 14.5cm]{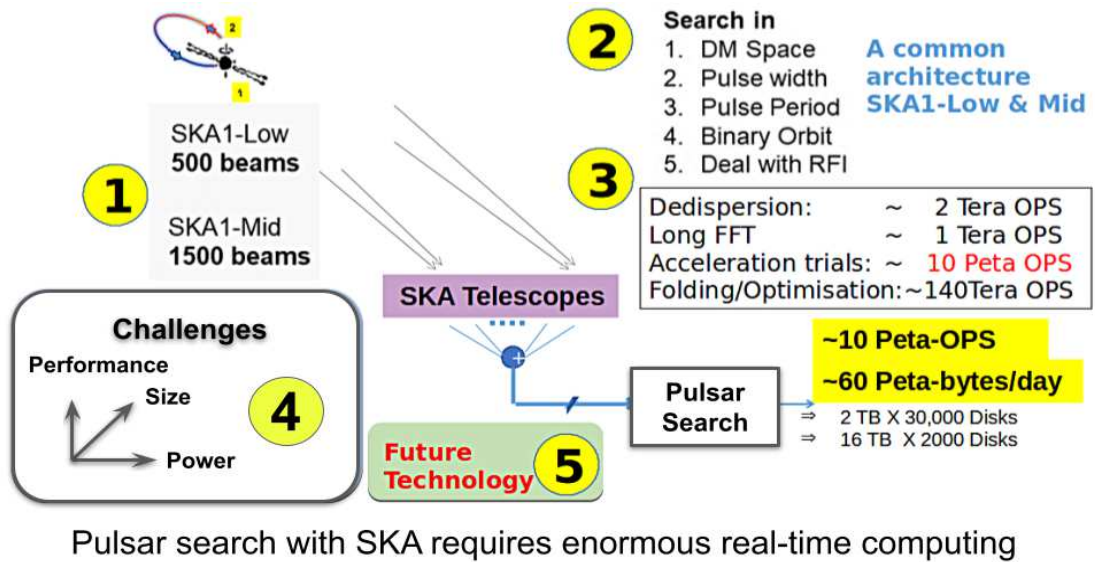}
        \caption{ SKA Pulsar Search: 
        1) observations carried out over 100's of beams,
        2) many parameter spaces are examined, 
        3) associated computation is enormous, 
        4) optimisation across power, performance, etc. is desirable, and
        5) prototypes to be implemented with future technology.\\}
        \label{fig:pss}
\end{figure*}

Transient events such as fast radio bursts (FRBs) and pulsars occur over a multi-dimensional parameter space, which may include: distance (dispersion measure space), pulse width, pulse period and binary orbit. Radio frequency interference mitigation is also required for all data streams. Search is also simultaneously carried out over several hundreds of voltage beams across large bandwidths. Hence the associated data volumes and compute complexity are enormous, amounting to 60 Peta-bytes/day and 10 Peta operations per second \citep{levin}. Such large amounts of data also mean that they cannot be easily transported or stored to process the data later offline. These aspects make the SKA transient search pipeline, depicted in Fig.~\ref{fig:pss}, compute-intensive, requiring specialized algorithms and high-performance computing to process the data in real-time.  

In recent years, there has been a trend towards offloading compute-intensive workloads to specialized devices referred to as “accelerators”. Accelerators have improved the platform performance in crucial computations. Notable changes contributing to this trend are Moore’s law and the accelerator's increasing programmability \citep{accelpcie, tan}.

The pulsar search team in the SKA have been investigating PCIe (Peripheral Component Interconnect Express) based FPGA \textbf{(Field Programmable Gate Array)} accelerators since the early commercial availability of these accelerators (section \ref{pulsarserach})~\citep{emit2015}.  We can now observe an increased use of FPGA-based accelerators for modern radio telescope applications (section \ref{skacase} \& \ref{sec:OngoingAndFutureOpportunities}).  

This paper focuses on FPGA-based acceleration of signal processing applications for radio telescopes with examples from the SKA, SKA-precursor and related application areas. 

The paper provides an outline of radio telescope signal processing in (section \ref{sec:radioTeleSignalProcessing}), compute complexity in radio-telescope signal processing application in (section \ref{sec:challRadioTeleSignalProcessing}), use of accelerators to simplify the compute problems in (section \ref{sec:accelLand}),  about the emerging FPGA-based accelerators in  (section \ref{sec:fpgaAccel}), their architecture in  (section \ref{sec:fpgaAccelArch}), configuration/programming model  in  (section \ref{sec:fpgaAccelReconfig}), code development tools in  (section \ref{sec:fpgaAccelCodeDev}), two major case studies in SKA, where significant investigation of the FPGA-based acceleration has been performed in  (section \ref{sec:fpgaAccelSkaCase}), about the ongoing and upcoming applications involving GMRT in (section \ref{sec:fpgaAccelGMRT}), on the use of ML suits with FPGA in (sections \ref{sec:fpgaAccelRandForest} \& \ref{sec:fpgaAccelMaskRcnn}), and about the prospective FPGA platforms and emerging open-source tools in (sections \ref{sec:fpgaAccelEmerging} \& \ref{sec:fpgaAccelRfsoc}). Then the paper concludes by capturing the summary of this review.

\section{Background}
\subsection{\textbf{ Radio telescope signal processing}}
\label{sec:radioTeleSignalProcessing}
\hfill

 Modern radio telescopes are very sensitive and versatile. There is always a need to employ many innovative approaches to best utilize the capabilities of these telescopes.

Incorporating innovative algorithms and real-time signal processing in a modern radio telescope is inevitable. In the past, such improvements needed expensive solutions involving custom built digital integrated solutions. Recent advances in digital technology have now made it possible to address these problems by replacing rigid, custom-made processing systems with reconfigurable/re-programmable digital systems enabled by modern technology.

From the computing point of view, one can abstract the various signal processing sections of a radio telescope into the following blocks:
\begin{enumerate}
    \setlength{\itemsep}{2pt}  
    \setlength{\itemindent}{10pt}  
        \item Sensor  
        \item Signal conditioner  
        \item Analog to digital converter
        \item Real-time processing / recording and
        \item Off-line processing
\end{enumerate}

A sensor converts the electromagnetic signals arriving from celestial sources into analogous electrical signals. Typically these are one or more radio antennas that act as a sensor (for eg., 197 antennas in SKA mid and 1,32,072 antennas in SKA Low). The signal conditioner amplifies electrical signals and passes selected frequency bands for digitization by an analog to digital converter (ADC). Digitized data from the ADC is processed either in real-time on special purpose hardware or recorded for offline processing.

Conventionally, each class of scientific motivations necessitated an independent combination of analog and digital hardware to be built for a radio telescope \citep{crs1}. Complex analog processing was also needed in the past from limitations of the digital systems to handle large bandwidths ranging from tens to hundreds of MHz. However,  recent advances in digital technology make it meaningful to substitute many such analog signal processing with programmable/configurable digital systems. Considerations for modularity and scalability play an essential role in realizing a common platform that can be reused across multiple applications \citep{crs2}. 

From a computational point of view, we can further summarize the various functional signal processing blocks for a synthesis interferometer radio telescope such as the SKA. Most of the standard operations can be formulated combining two specific operations which are correlation and beam forming. A simplified description of these steps are given below: 


\begin{align}
       \textbf{Digitization:} \hspace{4.5em} x[n] = x_c(nt)
\end{align}
where n=0,1,2,3, etc are the sample numbers, t the sampling period, \(x_c(t)\) the continuous-time signal.
\begin{align}
       \textbf{Delay Compensation:} \quad x_d[n] = x[n-n_d]\cdot e^{j.\omega.\delta_d.t}
\end{align}
where, \(n_d\) is a positive delay amount applied for delay correction. \(\delta_d\) is the fractional delay.
\begin{align}
       \textbf{Cross-correlation:} \hspace{1.0em} x_{1,2}[n] = \sum_{m=-(\frac{N}{2}-1)}^{(\frac{N}{2}-1)} x_1[m]\cdot x_2[m-n]
\end{align}
where, \(x_1\) and \(x_2\) are the outputs from two different antennas, and N being correlation length. In the case of the correlators that are implemented after frequency channelisation (filter bank) stage, the cross correlation  will be carried out by multiplying respective frequency channels between  all antenna pairs and then averaging them. This kind of correlation will not have the time lag (m=0 always) that appears in equation (3) and is referred to as FX (Fourier transform followed by multiplication) type.

\begin{align}
       \textbf{Beam Formation:} \hspace{1.0em} x_{pa}[n] = \sum_{m=1}^{q} x_m[n-n_d^m]\cdot e^{j.\omega.\delta_d^m.t}
\end{align}
where q is the number of antennas, and \(n_d^m\) are the delays applied for the \(m^{th}\) antenna.  \(\delta_d^m\) is the fractional delay for \(m^{th}\) antenna.

A detailed description needs to include dynamic delay and fringe updates, which are not relevant for brevity; the reader may refer to \citep{thompson}.

 \begin{figure}[H]
    \centering
    \includegraphics[width=0.35\textwidth]{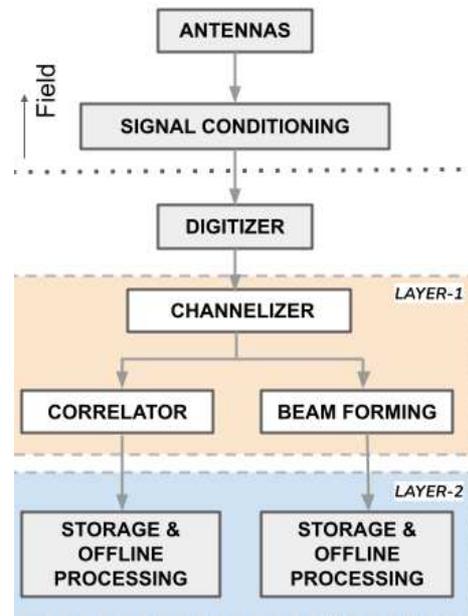}
        \caption{Digital signal processing in a radio telescope happens in two-layers: LAYER-1 \& LAYER-2. In SKA and other upcoming telescopes, the LAYER-2 applications are executed in real-time using powerful accelerators. }
        \label{fig:twolayers}
 \end{figure} 

With the continuously increasing processing power of relatively less expensive networks of computers and powerful modern electronics, it is also helpful to visualize the real-time processing into two layers from a specific implementation point of view. As shown in Fig.~\ref{fig:twolayers}, the first layer (layer-1) of processing typically involves a deterministic sequence of latency-critical operations such as spectral decomposition, correlation, phasing for beam formation. In contrast, the second layer (layer-2) of processing covers a wide range of application-specific processing and data recording. From the signal processing speed performance points of view, full custom VLSI-based or gate arrays are convenient for processing at layer-1. At the same time, a network of general-purpose workstations equipped with computing accelerators are more suitable for processing at layer-2.

In a typical signal processing backend at a present-day radio telescope (either a single antenna or an array of antennas), most of the processing is performed by digitally employing a combination of custom-VLSI, processors and field-programmable arrays. Most high-speed signal processing is achieved through VLSI devices, but is increasingly being implemented with FPGAs. For instance, initially, full-custom chips have been used as main computing elements for the first stage of real-time processing in the Very Large Array (VLA) in the US, Westerbork Synthesis Telescope (WSRT) in the Netherlands, the Australia Telescope Compact Array (ATCA), Very Long Baseline Array (VLBA), and the Giant Metre Wave Radio Telescope (GMRT) in India. Newer telescopes such as the Murchison Widefield Array (MWA) in Australia, CHIME in Canada, MeerKAT telescope in South Africa, the upgraded GMRT, and the proposed upgrades for ORT \& GMRT \citep{network,scalable} employed/recommend FPGA for delivering the most complex high-speed digital signal processing building blocks in their signal chains.   
 
Thus, novel developments have been occurring in recent years, especially in the SKA for layer-2 processing. \textbf{This paper focuses on layer-2 computing for the SKA and highlights the possibility of using FPGA-based accelerators for layer-1 processing in the SKA-pathfinders\footnote{\url{https://www.skatelescope.org/precursors-pathfinders-design-studies/}} as well as for other upcoming applications}. 

\subsection{\textbf{ Challenges in processing radio telescope signals}}
\label{sec:challRadioTeleSignalProcessing}
\hfill

Various computing challenges are to be dealt with in radio telescope signal processing. Complexities differ from one kind of application to the other. In some cases, a large amount of data needs to be handled. In other instances, the data volume could be lower, but compute complexities could be significant, requiring high precision arithmetic to avoid numerical errors and faster executions to meet the real-time constraints.\\

Typical challenges  faced in a modern radio telescope signal processing can be listed as given below:

\begin{itemize}
    \setlength{\itemsep}{2pt}  
    \setlength{\itemindent}{10pt}  
        \item Very high speed data
        \item Huge volumes of data \label{datasize}
        \item Many input/output streams \label{numio}
        \item A large variety of algorithms
        \item Varying complexity with algorithms
        \item Complexity with data access
        \item Spatial dependency of the data
        \item Time available for computation
        \item Latency in parallel processing 
        \item Precision considerations
        \item Performance per watt \label{pperf}
        \item EMI/EMC compliance
        \item Thermal issues
\end{itemize}

These complexities restrict the hardware implementation choices of the algorithms. Among this list, the constraints arising from a)  \textit{Huge volumes of data}, b) \textit{Many input/output streams}, and c) \textit{performance per watt} seem to be influencing the {memory, input/output architecture and technology choices} of new generation compute accelerators. The subsections  \ref{memlimit},  \ref{iolimit} and  \ref{pwrlimit} discuss these three aspects respectively, and outline some choices available to deal with them. 

\subsubsection{\textbf{Memory Performance:}}  \label{memlimit}
\hfill
Memory is an essential component in high-performance computing. In many signal processing applications, data access becomes a bottleneck limiting the computational capabilities of an accelerator. A proper choice of the memory capacity and interface is most important for high-performance computing.
\begin{figure}[H]
    \centering
    \includegraphics[width=0.2\textwidth]{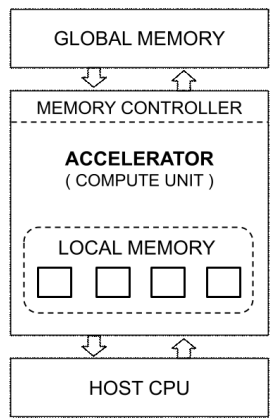}
        \caption{ The configuration of Global memory and Local memory in an accelerator. The global memories are large in capacities whereas the local memories are smaller but have higher bandwidth.  }
        \label{fig:Memory}
\end{figure} 
 Memory requirements for a real-time performance come in two categories: 1) local memory: a large number of small capacity memories and 2) global memory: a few banks of high capacity memories as shown in Fig.~\ref{fig:Memory}. It is also desirable to have the small capacity memories to facilitate quicker access (or low latency). The local memories are realized by using an accelerator's registers and internal buffers. Although it is advantageous to have global memories with low latency, usually they come with high latency due to implementation architectures \citep{MEM}. 
 
 In recent years, we have seen new technology memory devices in the market that can help implement global memories. At the time of this study, most accelerator boards support DDR4 memories, whilst some vendors are proposing integrated High Bandwidth Memories (HBM) as tiles to the FPGA cores and GDDR6 SDRAM in their cards. HBM memories appear  attractive due to the large bandwidth that they can offer. On the other hand, DDR4 memories seem better in terms of cost and power performance due to the large consumer market and reduced core voltages used in DDR4 chips \citep{Villegas2018,McKee}.

\subsubsection{\textbf {Input-Output Performance:}} 
\label{iolimit}
Correlators of the radio telescope arrays need to exchange large amounts of data across the processing elements rapidly and would require a very efficient input-output speed performance. A low latency interface to the host and external systems becomes essential in such accelerated processing.

The PCIe and Gigabit Ethernet were potential interface technologies to facilitate such data movements. During this study, several accelerator cards use PCIe based bus. Besides PCIe, some accelerators also have Ethernet capabilities and dedicated interconnect buses. We also observe that both PCIe and Ethernet are migrating to higher and higher bit rate capabilities. During this study, 16 Gbps capable host interfaces and 100 Gbps inter-connectivity seem state of the art \citep{parson, network, crs3, crs5}.     

\subsubsection{\textbf{Performance per watt:}}  \label{pwrlimit} 
 
Various factors such as clock speed, semiconductor technology, algorithmic efficiency, effective utilization of computing resources, etc., affect the performance efficiency achieved per unit of power consumed in an accelerator.  An increase in the accelerator clock speed delivers a higher number of operations per second (OPS). However, the semiconductor power dissipation also increases exponentially with higher clock speed. Thus, the higher the clock rate, the higher the amount of power dissipated on the accelerator device. In this section we will look at the complexity associated with thermal dissipation.  

With the semiconductor \href{https://encyclopedia2.thefreedictionary.com/feature+size}{feature-size}\footnote{\url{https://encyclopedia2.thefreedictionary.com/feature+size}} shrinking to 10~nm and 7~nm, many transistors sit around a small place. There is starvation for heat dissipation paths due to the increased transistor density. Hence with the shrinking of the semiconductors, substantial thermal issues arise. They are now being dealt with using conventional solutions. A variety of cooling options: fan-based, liquid cooling etc., are being tested on modern accelerators, but compact solutions are still on the horizon. \href{https://en.wikipedia.org/wiki/Thermal_runaway}{Thermal runaway} \footnote{\url{https://en.wikipedia.org/wiki/Thermal_runaway}} of the semiconductor is still an issue, as higher rates of computation cause more heat dissipation. Typically, the accelerator integrated chip semiconductor junction temperature needs to be maintained below 100\textdegree C for reliable performance and longevity. 

\begin{figure}[H]
    \centering
    \includegraphics[width=0.4\textwidth]{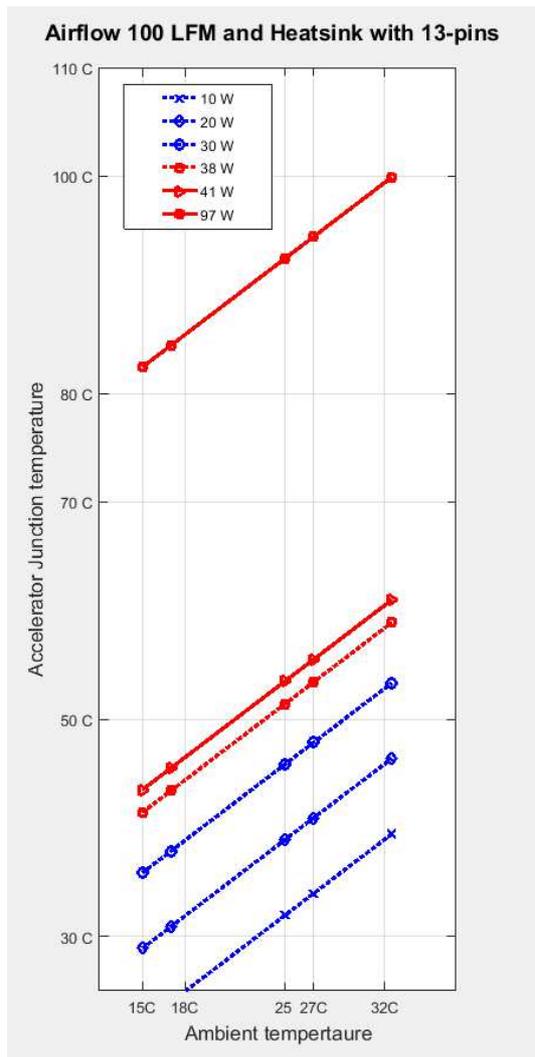}
       \caption{ Plot illustrates the effect of accelerator power and ambient temperature on the  junction temperature that needs to be maintained below 100  $^{\circ}$C for a reliable operation. Results obtained through modeling an use-case (section \ref{pwrlimit}).}
       \label{fig:thetaj}
\end{figure}

  To provide a feel for the complexity with thermal aspects, we present in plot Fig.~\ref{fig:thetaj}  an illustration of the effect of on-chip power dissipation on semiconductor junction temperature. This example assumes specific nominal temperature coefficients for the accelerator package, 13-pin type heat-sink and 100 linear-feet per minute (LFM) airflow values, while the accelerator power dissipation is varied from 10 to 97 W of power. At the higher side of 97 W power, the ambient temperature is to be maintained well below 32 $^{\circ}$C (or the airflow is to be increased) to keep the junction temperature below 100 $^{\circ}$C for the reliable operation of accelerator hardware. Details of the calculation are available in \href{https://github.com/tprabu2000/Power-Performance}{archive}\footnote{\url{https://github.com/tprabu2000/Power-Performance}}. 
 Accelerators and server motherboards incorporate onboard power and thermal management circuits to make real-time measurements on critical components such as the memory, FPGA and GPU (Graphics Processing Unit). Usually, making direct measurements on such devices is difficult with conventional tools. These onboard power management tools are non-invasive; hence monitoring the power and thermal aspects on fielded accelerators becomes easy.

In the subsequent section, we present details of specific accelerator technologies developed in the recent past addressing the computing challenges.  
\section{FPGA-based accelerated computing}

\subsection{\textbf{Accelerator Landscape}}
\label{sec:accelLand}
\hfill

Accelerators are devices optimized to improve performance of computing systems, such as CPUs. Optimization is achieved by implementing parallel execution paths or using hardware built specifically for an algorithm. 

In the recent past, manycore and multicore processors have been used as accelerators in certain areas of computing. To provide a feel for the variety of architectures developed to accelerate computing, we present here salient details of Plessey, DSP,  Transputer, GPU, TILERA, KalRay, Xeon Phi, TPU and FPGA.

\begin{figure}[H]
    \centering
   \includegraphics[width=0.45\textwidth]{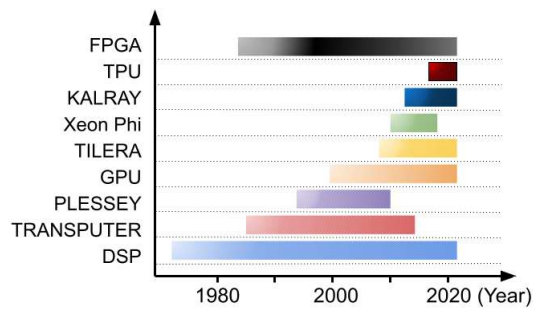}
        \caption{ Timeline of active publications observed for different accelerators.  }
        \label{fig:aceltimeline}
\end{figure} 

Plessey system is an operational computer which is designed on the basis of capability-based addressing. Plessey 250 had a multiprocessing architecture allowing connection of up to eight processors with shared memory across all \citep{PLES}. The memory was divided into segments for each of these processors, such that one processor did not affect the computation of other processors. Each instruction was 24 bits and input/output devices were controlled by special device registers that existed in physical address space.
 
DSP processors are used to handle large amounts of complex mathematical operations quickly. The first ever DSP processor built was for synthesizing speech by Texas Instruments \citep{DSP}. Over the years, with growing technology, DSP processors could do more mathematically intense scientific calculations \citep{psr}.  The ADSP-SC57x and ADSP-2157x are latest DSP multicore processors from Analog Devices, which provide high performance with less power consumption and deliver greater than 6 giga-floating-point operations per second using two enhanced SHARC+ cores and advanced DSP accelerators. 

Transputer was one of the first general purpose microprocessors which was specifically meant for computing in parallel \citep{TRANS}. Transputers could connect to other transputers for communication. Around 1980, Inmos had developed transputers consisting of a small processor, 4 KB of RAM, 4 communication links (one link provided 2 channels in both directions) providing connection via 2-wire to another transputer and an external memory interface. These transputers provided high performance which could be used for many applications.

A graphics manipulation and output device known as GPU became a very attractive compute accelerator in modern high-performance computing. It is mainly due to the availability of several 100s of computing cores and several thousands of threads in GPUs. GPUs can yield performances in 100s of Gigaflops to Teraflops, with single and double-precision floating-point processing capabilities. GPUs can be programmed using CUDA, Python, C and OpenCL languages \citep{andre, peng}.   

TILERA is a multicore ($\sim$288) processor connected by an onboard network \citep{TIL}. The I/O connectivity of the cores is achieved via a terabit switch and it also has a high-performance transaction port for chip-to-chip or FPGA interconnect. Standard C/C++ and Java programming support are available. High-end TILE-Gx devices can address up to 1 TB of memory. They are typically implemented on PCIe cards with DDR memories and multiple 10 Gb Ethernet ports. 

Kalray is a manycore processor with 64 to 1024 core processors, composed of a cluster array connected through a high-speed network-on-chip implemented on PCIe cards. Each cluster contains 16 processing cores, a system core and a shared memory. High-speed interfaces are achieved through Ethernet. C/C++/FORTRAN-based parallel programming solutions are available for development \citep{KAL}.

Xeon Phi is a manycore X86-based processor, with typically many ($\sim$72) cores in a single chip \citep{Xeon}. Xeon Phi architecture allowed running applications that generally ran on any X86 CPUs. A Xeon Phi coprocessor could host an operating system, be fully IP addressable and support standards such as message passing interfaces (MPI). Typically, Xeon Phi was available as a PCIe card powered through PCIe and auxiliary power connectors. 

Tensor Processing Unit (TPU) is a PCIe-based accelerator, developed by Google for accelerating Tensor-flow machine learning / neural networks applications. Edge TPUs are one of the latest models which can perform 4 Tera operations at 2 W of electric power \citep{tpu}.

\begin{figure}[H]
    \centering
    \includegraphics[width=0.5\textwidth]{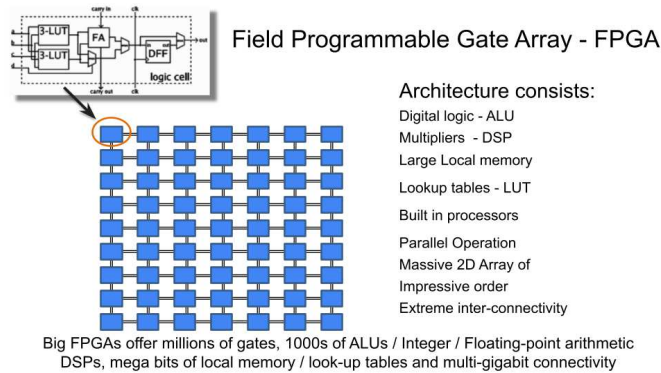}
        \caption{ Field Programmable Gate Arrays (FPGAs) are designed using fine-grained, massively parallel, arrays of digital logic with architecture suitable to execute computations in parallel. }
        \label{fig:fpga}
\end{figure} 

The introduction of Field Programmable Gate Arrays (FPGA) revolutionized traditional approaches to digital circuit design \citep{Alfke2011, Kingdon1992, Lytle1993, Stephen2015}. Radical changes introduced to the digital design approach can be compared to the impacts after the programmable loom of Jacquard [early 1800s] and after the universal machine of Turing [1937]. 
FPGAs are designed using fine-grained, massively parallel arrays of digital logic with architecture suitable to execute computations in parallel (see Fig.~\ref{fig:fpga}).  

While these accelerators have been used to solve various compute-intensive acceleration problems, for example, Transputers have been used for early commissioning of major radio telescopes \citep{Swarup}, Plessey PDSP16510 FFT accelerator have been experimented with the RRI-Millimeter wave radio telescope, 
they stay only as good architectural references. On the other hand, DSP and GPU-based accelerations applications target large user communities. Consequently, their development ecosystem is growing to much higher maturity levels. New generation accelerators such as those based on TPU are entering to the machine learning applications; FPGAs-based accelerators are entering compute areas traditionally occupied by DSPs and GPUs. While FPGAs in the early days offered handy solutions for simple digital circuitry, modern FPGAs are becoming very powerful and promise energy-efficient solutions for computation-intensive problems.\citep{alterahpc}. In subsequent discussions, we will focus on FPGA-based accelerators.

Present day FPGAs have large numbers of digital signal processing modules (DSP) capable of performing high-speed multiplication, addition, accumulation, fixed-point and floating-point operations, large amounts of high-bandwidth distributed memories (local memories), mature FPGA development tools and electronic design automation processes. This, along with radical approaches to adapting to host-device architectures help realize purpose-built implementations to accelerate complex algorithms.

The use of FPGAs involves programming it to represent a circuit that performs the algorithm of interest. Suitable mechanisms are required to describe the algorithm's serial and parallel (concurrent) operations to be implemented and a provision to analyze and synthesize the circuit operation. A variety of reasons, including PLD fabrication processes used in FPGAs, have improved their bare performance today. New FPGAs now promise more than 10 Tera-FLOPS of performance at 1 Tera-FLOPS/watt \citep{tan}. We also see that data centers tackling streaming data and Big Data are moving to use FPGA accelerators to handle these next-generation data processing challenges.  

In the last decade, FPGAs have entered into high-performance computing due to radical changes to software development approaches and advanced semiconductor fabrication processes offering significant savings on power with FPGA-based acceleration. Xilinx (now part of AMD) and Altera (now part of Intel) are the major manufacturers of high-performance FPGAs today. Both have suitable FPGA platforms for data center, artificial intelligence (AI) / machine learning (ML), signal processing and imaging applications. 
 
Multiple vendors are also developing PCIe-based accelerator cards using Xilinx and Intel FPGAs. Programming support for such platforms is available via both HDL and high-level synthesis tools. Further, the host CPU interface via regular PCIe OS driver software is also evolving well\citep{Vesper}.

 \subsection{\textbf{FPGA based Accelerators}}
\label{sec:fpgaAccel}
\hfill

\begin{figure}[H]
    \centering
    \includegraphics[width=0.45\textwidth]{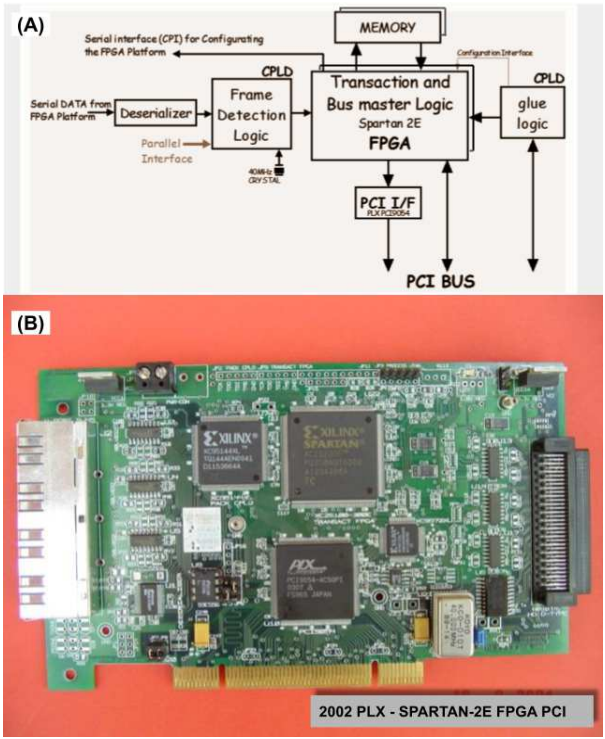}
        \caption{ (A) Layout of an early, in-house built 2002 PLX SPARTAN-2E FPGA PCI accelerator. It has dual SPARTN-2E FPGAs, dual onboard SRAM memory, 380 Mbps serial and 16-bit parallel interfaces for the I/O. Host CPU configures the FPGAs through PCIe. An onboard CPLD manages configuration transactions of the FPGAs. The PCIe interface is through a dedicated PLX ASIC (Modern FPGAs have hardened PCIe IP blocks). (B) Picture of the FPGA board. The second FPGA and memory modules are located on the rear side.}
        \label{fig:plx}
\end{figure}

The choice of using FPGAs as accelerators has existed for more than two decades \citep{yogi,ort}. Although many of these early attempts were implemented to solve dedicated problems, they remained experimental prototypes for a long time, until large size FPGAs came into existence. Early FPGA accelerators (although seldom referred to as accelerators) required separate interface chips for connecting to the host via PCI bus. In order to show the contrast of these differences we present details of an in-house developed 2002 FPGA board in Fig.~\ref{fig:plx}.    

About ten years down the lane in semiconductor history, FPGA capacities increased many folds, and FPGA based PCIe accelerators came into existence as commercial off the shelf (COTS) products in the market. In addition, tools for programming FPGAs also saw significant developments. For a long time, hardware description languages (HDL) used to be the only means for developing reliable codes for FPGAs. Subsequent development of automatic code generators based on Matlab and National Instrument tools, platform-independent languages and high-level language-based tools for FPGAs made them accessible to programmers with no prior HDL development experience. In addition, FPGAs promised a heavy reduction in energy consumption compared to traditional platforms. These aspects proved to be attractive for data centers and companies such as Google and Microsoft started adopting FPGA accelerators in their data centers. Around this time, CPU - GPU vendors also started manufacturing FPGAs and introduced standard development flows to work across CPU, GPU and FPGA. These events further enhanced the FPGA accelerator market and broadened its usability. 

Most modern FPGA accelerators are beginning to be available in the COTS market. We observe the AMD and Intel accelerators aim to support high-performance computing for various applications: Big Data, streaming signal processing, power-efficient computing, and machine learning applications. A wealth of high-level programming tools, including C/C++, OpenCL and OneAPI, are becoming available for the FPGA accelerator code developments.

\begin{figure}[H]
    \centering
    \includegraphics[width=0.45\textwidth]{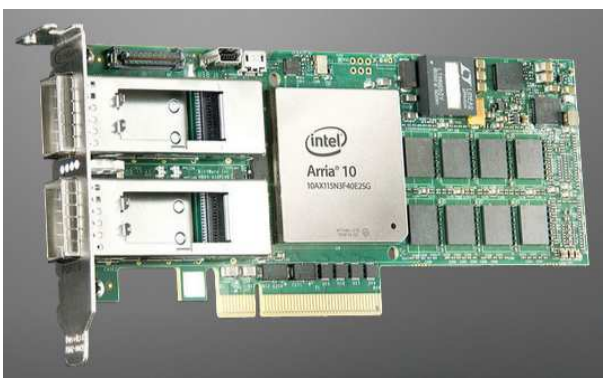}
        \caption{ COTS FPGA accelerator BittWare A10PL4 } 
        \label{fig:arria10}
\end{figure}

An example of a recent COTS FPGA accelerator\footnote{\url{https://www.bittware.com/fpga/a10pl4/}} is shown in Fig.~\ref{fig:arria10}. A general layout of a COTS FPGA accelerator and firmware is shown in Fig.~\ref{fig:firmware}. The FPGA connects to the host CPU through a PCIe bus. There will be a fixed number (two or four nominally) of DDR memory banks for global memory, which connect to the FPGA through dedicated  (64 or 72 bits wide) memory bus interfaces. A few peripherals, such as Ethernet etc., will also have connections to the FPGA.

\begin{figure}[H]
    \centering
    \includegraphics[width=0.5\textwidth]{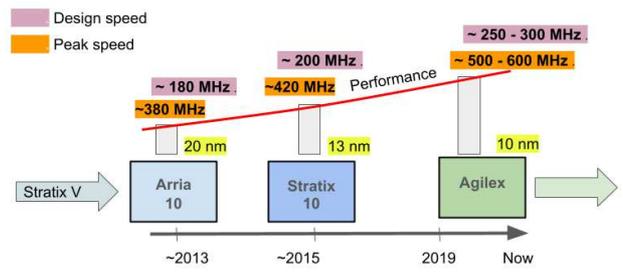}
        \caption{ Clocking performance trends of FPGAs over the years.}
        \label{fig:speed}
\end{figure}

 Bare FPGA performances are increasing over time with miniature feature-sizes being adapted for fabricating the latest FPGAs. Fig.~\ref{fig:speed} shows the maximum clock speed trends observed with the latest FPGAs. 

\begin{figure}[H]
    \centering
    \includegraphics[width=0.5\textwidth]{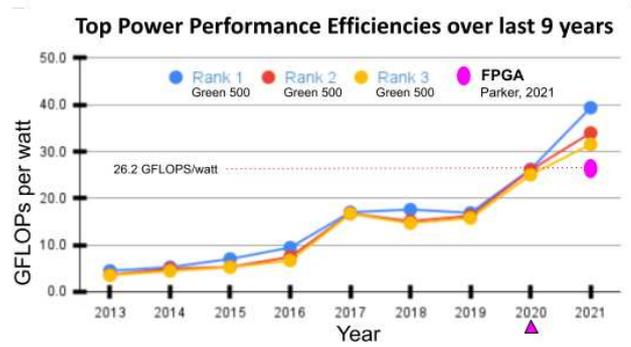}
         \caption{ Best three power performances (Rank1:~${\sim}{40}$ GFLOP/Watt, Rank2:~${\sim}{34}$ GFLOP/Watt \& Rank3:~${\sim}{30}$ GFLOP/Watt in 2021) reported by Green500 over the last 10 years is shown in the figure. It can be observed that FPGA accelerator's (2020) power performance (26.2 GFLOPS/Watt) is closely matching these 2020 top power performances.}
        \label{fig:gfloppw}
\end{figure}

Power performances reported by  Green-500\footnote{Source: \url{ https://www.top500.org/resources} } computing show improvements from 5 to 40 GFLOPS/watt over the last decade. This trend is shown in Fig~\ref{fig:gfloppw}. Power efficiencies of the modern FPGAs are also closely tracking these top performances. FPGA accelerator power performance (26 GFLOPS/W) reported by \citep{Parker2021} is also shown in the plot.

Next section provides an overview of COTS FPGA accelerators and aspects of programming the FPGA accelerators.  
\pagebreak

\subsection{\textbf{FPGA Accelerator Architecture}}
\label{sec:fpgaAccelArch}
\hfill

\begin{figure}[H]
    \centering
    \includegraphics[width=0.45\textwidth]{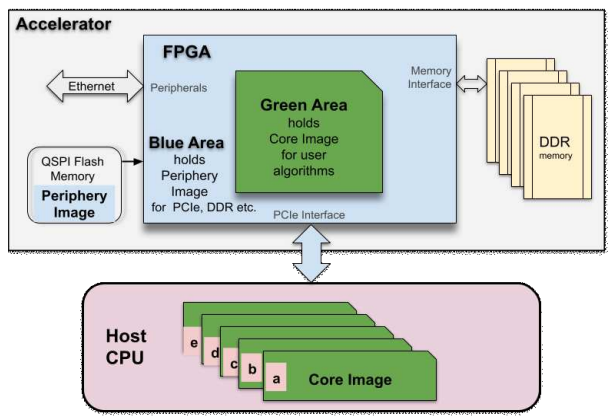}
        \caption{Accelerator FPGA green area resources runs the user application. Host CPU dynamically configures this area from a set of available images, through the PCIe bus. Configuration of the blue area is fixed and images are supplied by the vendor. This image provides interfaces to memory, Ethernet etc. It is loaded automatically from a flash memory on power-up. }
        \label{fig:firmware}
\end{figure}  
 FPGA accelerator have certain fixed functionalities, such as: 
 
  \begin{enumerate}
    \setlength{\itemsep}{2pt}  
    \setlength{\itemindent}{10pt}  
        \item PCIe interface to the host PC 
        \item Configuration scheme  
        \item Interface to DDR memory, and
        \item Peripheral interfaces: Ethernet etc.
\end{enumerate}

The central area of the FPGA, shown with green color in Fig.~\ref{fig:firmware}, will be reconfigured for each new application. The periphery areas shown in blue color holds a static design, and they need not change for different applications. 

Memories and bus interfaces used in the FPGA accelerator board decide how the blue area is configured. However, this part of the configuration remains static for a given accelerator. Hence, accelerator vendors provide needed libraries or example codes to configure the \emph{blue area}. In comparison, the \emph{green area} is maximally constrained only by the algorithm complexity and the implementation choices (for example, number of memory banks needed etc.) considered for the board. 

\subsection {\textbf{FPGA Accelerator Reconfiguration }}  
\label{sec:fpgaAccelReconfig}
\hfill

Accelerators typically execute specific codes or kernels for different application instances. In case of FPGA accelerators, kernels are precompiled to form what is known as a bit file. They are a stream of bits, typically up to a few tens of megabytes (a fixed size for a given type of FPGA), that will be loaded into the FPGA to define the hardware architecture for acceleration. This loading of the bit file is also known as the FPGA configuration process. Most FPGAs support two protocols for the configuration process: at first, the FPGA initiates and reads an external bit file by itself. In slave mode, an external device, typically a CPU, acts as a master and writes the bit file. PC based accelerators use the latter described slave mode, and the host CPU configures the FPGA with different bit files needed for the application. Although multiple physical interfaces are available for configuring FPGAs, present accelerators rely on a PCIe-based FPGA configuration, also known as configuration via protocol (CvP). 
An illustration of the configuration flow is also shown in Fig.~\ref{fig:firmware}. FPGAs use hard-IPs (intellectual property libraries that are integrated in the FPGA) to implement the PCI interface. In order to use CvP, the PCIe interfaces of the FPGA are to be brought up first and hence the configuration is conveniently divided into two parts: 1)  Peripheral update, and 2) Core update.\\

The peripheral update configuration takes place automatically as soon as the host CPU is powered up. It is a mandatory housekeeping step executed only once. This step brings up the FPGA periphery containing the PCIe links very quickly, within 100 ms 
When the PCIe link is up, the accelerator can now connect to the host CPU motherboard in time and reveal its presence to the host OS drivers. The configuration of the peripheral areas are static (for a given tool flow version), and user applications need not modify them. Also, a comparatively small portion of information from a bit file is required to update the peripheral section. Hence, the peripheral update configuration information, also known as a periphery image, is split separately from a bit file (or pre-generated in the case of OpenCL) kept in an onboard non-volatile Quad SPI flash memory. 
Thus the accelerator board automatically performs peripheral updates from the flash memory upon power-up.  This first part of the configuration achieves physical connectivity between the accelerator and the host. 

Subsequently, user programs invoke core updates. The core update configures the green part of the FPGA shown in Fig.~\ref{fig:firmware}. The green parts of the FPGA contain a wealth of reconfigurable hardware resources such as arithmetic, logic,  local memory, counters etc. These resources get rewired specific to the architectural needs of different acceleration applications. The bit file loaded for this update is a core image created by the FPGA synthesis tool (for example Quartus 
for Intel FPGAs, and Vivado for Xilinx FPGAs). Various core image updates can use the same peripheral image. The user can have many core images, but they all need partitions with identical periphery images. Usually, this is not a concern, as development tools take care of such partitioning tasks. Once the core update is complete, the accelerator can run the user applications.  Thus the accelerator board core update is through user applications from host and multiple core images can be stored in a host.  This first part of the configuration achieves connectivity between host applications and kernels in the accelerator. 

\subsection {\textbf{HDL and HDL generators}}
\label{sec:fpgaAccelCodeDev}
\hfill

Hardware Description Languages (HDL) are specialized languages used to express the structure and behavior of digital logic circuits. These circuit descriptions can be simulated using software packages that let you check the values of signals inside your design and enable testing of designs before they are turned into hardware. They can also be synthesized for both FPGA and ASIC targets. Synthesizable HDL designs are typically written at the Register-Transfer Level (RTL), a digital design abstraction used to describe circuits based on how data flows between registers and the logic that operates on this data.

VHDL and Verilog are two of the most popular HDLs. VHDL stands for VHSIC (Very High Speed Integrated Circuit) Hardware Description Language and was developed in 1980 by the US Department of Defence. It is a strongly typed language with a non-C like syntax, and its highly verbose nature allows designers to describe designs with comparatively lower ambiguity. On the other hand, Verilog is a more compact and less wordy language with a C-like syntax, making it easier to understand for designers with knowledge of C. It is more akin to a modeling language, allowing for designs to be written faster but can potentially have non-deterministic behavior.

Newer HDLs like Chisel, Bluespec Verilog, SpinalHDL, Migen etc., have also been developed to improve design productivity. Chisel, for instance, is embedded in the Scala programming language and compiles to synthesizable Verilog. It brings object-oriented and functional programming paradigms and the Scala development ecosystem to hardware description. Its strong parameterization capabilities allow designers to write circuit generators, enabling greater re-usability of code. While being close to RTL, this raised abstraction level makes it easier to create complex designs. Other languages listed above also use various methods to achieve similar productivity gains compared to traditional HDLs.

FPGA toolchains typically support VHDL and Verilog-based designs. These tools provide various utilities, including simulating, synthesizing, applying constraints, place and route (placing logic elements and interconnecting them on the FPGA grid) and programming the FPGA. However, these tools may require accurate knowledge of the underlying hardware architecture of the FPGA, needing long design cycles to even reach the first-level implementation for complex circuits. This is compounded by the fact that HDLs can be hard to use and debug. FPGA designers may also face difficulties in exploiting resources of large FPGAs in reasonable time. Many such toolchains provide automated IP generator capabilities to alleviate this, allowing designers to generate large designs through code-generation scripts. 

One of the main advantages of implementing designs at RTL is that it offers fine-grained control to the developer. Although this works well for standalone FPGA platforms, designs that need to work on heterogeneous platforms, where FPGAs interact with each other or with a conventional processor environment, have to deal with much greater complexity.

\subsection{\textbf{High level language tools}}
\label{sec:HighLevelLanguageTools}
\hfill

Given the long design cycles when using HDLs for developing FPGA-based accelerators in HPC environments, non-verbose high level language tools have emerged as an attractive option. Such tools make acceleration especially accessible to software developers and provide a middle ground between hardware and software. This also means development is much faster and porting an application to a new platform is easier.

The Open Computing Language (OpenCL) framework, introduced in 2008, is one such platform. It utilizes application programming interfaces (APIs) to control and execute programs across heterogeneous platforms consisting of CPUs, GPUs, DSPs and FPGAs. The framework provides parallel computing using task-based and data-based parallelism and uses library layers and driver functions to hide complexities of hardware from the user. It also implements methods to schedule tasks and manage hardware resources such as processors, memory, data storage and I/O devices. Subsequent to its development, the two major FPGA vendors, Altera and Xilinx, announced OpenCL support in their HPC platform design flow. 

In a typical HPC system with a FPGA accelerator interfaced to a host CPU, OpenCL codes consist of host (CPU) side and device (FPGA) side codes. The code for the FPGA side is organized in kernels. Kernels are FPGA functions that can be invoked and sequenced from the host code. The host side code is written using the C99 style and uses a C++ compiler to produce the host side executable. The FPGA side code is generally intermixed with C++ and vendor-supplied libraries and use FPGA synthesis tools to create the FPGA binary (or bit stream/bit file). The FPGA binary generation from OpenCL involves HDL generation, synthesis, optimization, fitting, routing and timing optimization. This process invokes the FPGA tools in the background and takes a few hours (for complex designs) to generate the binary. Once the FPGA binaries are produced, they are deployed during the execution of the host code. OpenCL designs implement multi-threading by utilizing the deep-pipeline stages in the design. FPGA binary kernels are typically designed to follow a FIFO (First In First Out) based data distribution. The host program is responsible for scheduling, queuing and supplying data for the kernels in the FPGA, and the host-device codes work hand in hand.

OneAPI is a very recent open standard developed by Intel, that supports cross platform accelerator program development, including GPUs, FPGAs and AI accelerators, through a single programming environment. Essentially, it has most of the features of OpenCL to work in such a heterogeneous accelerator environment, but with a more uniform interconnect mechanism. OneAPI is based on SYCL (pronounced ‘sickle’),  a cross-platform abstraction layer that enables code development for heterogeneous accelerators to be written using a standard ISO C++ with the host and kernel code for an application contained in the same source file. Data Parallel C++ (DPC++) is the core programming language utilized by OneAPI. It uses a data parallel programming model, where the data is distributed across a set of processing elements, and each processing element operates on the data in parallel, allowing developers to scale parallelism with data size. Parallelism can be achieved by explicitly parallelizing an algorithm, or by using a parallel implementation from a OneAPI library. OneAPI makes it possible to target CPU, GPU, or FPGA with a single implementation by providing consistent interfaces. 


OneAPI also provides tools for working with deep learning models. This library contains basic building blocks such as convolution, matrix multiplication, pooling, batch normalization, activation functions, recurrent neural network (RNN) cells, and long short-term memory (LSTM) cells. 

While vastly boosting productivity and portability, the use of high-level languages on FPGA also faces some issues. One caveat is that such tools obfuscate the underlying architecture of the design, and developers have only indirect control over it. The loss of verbosity can make it difficult to steer the tool towards a good architecture, and platform-specific tuning is often necessary \citep{ZohouriHighPerformanceComputing2018, veenboerRadioAstronomyImaging2019, GozPerformanceAndEnergy2020}. Thus, using such high-level language tools versus HDLs can be seen as a trade-off between dealing with design complexity and having fine-grained control over the design.


\subsection{\textbf{Development environment}}
\label{sec:DevelopmentEnvironment}
\hfill

Code development tools for FPGA accelerators used to be an expensive option. However, major FPGA accelerator developers like Intel and AMD offer attractive development environments. Intel operates a cloud-based HPC development environment known as DevCloud. DevCloud comes with pre-installed OpenCL and OneAPI tools and libraries. A limited (but normally extendable) complimentary access to this development site is available to developers. Similarly, AMD (Xilinx) FPGA accelerators and Xilinx software development tools are available as services on Amazon Web Service instances. These cloud-based code development environments offer full support of their most modern software. 

\section{Case Studies} \label{skacase}
\label{sec:fpgaAccelSkaCase}
\hfill

In this section, we present a review of FPGA acceleration use cases relevant to the SKA. 

\subsection {\textbf{Pulsar Search with SKA}} \label{pulsarserach}
\hfill

Significant scientific contributions have come since the discovery of the first pulsar in 1967. Pulsars are highly magnetized, rotating celestial objects with extreme physical properties, and about 2600 of these objects have been located so far. This count represents only a tiny fraction of the available pulsar population yet to be discovered.

 A powerful telescope, such as the upcoming SKA, can detect faint signals from many thousands of these objects. Finding more of these objects would facilitate systematic scientific investigation of their nature and hence is a critical ongoing effort in astronomy. Therefore an extensive pulsar search program has been planned with the upcoming Square Kilometre Array (SKA) telescope.     

 The search effort involves real-time processing of significant amounts of data from the SKA telescope for each observation. Real-time processing is mandated due to the mammoth volume of data generated by the SKA telescope during a search. It cannot be practically stored for any offline processing. High-performance computing (HPC) solutions are explored to cope with real-time processing. 

The computational complexity in the search application comes from several factors. First of all, pulsars are located at immensely large distances from us. They are rotating objects, sometimes in orbit with other objects. Hence, when pulsar signals arrive on Earth, they are extremely faint, dispersed, Doppler shifted, submerged in noise due to various backgrounds, and also get affected by terrestrial radio frequency interference (RFI). These factors modify signals from the pulsar and need to be dealt with during a search operation. Hence, pulsar search involves processing to account for:  a) a range of distances to these objects, b) a range of periods for their rotation, c) a range of accelerations corresponding to their orbital motions and d) attention to distinguish actual pulsar signals  from RFI. To improve the measurements' efficiency, the telescope signals are quantized into multi-bit, the observations use a wide band of frequencies, and the sky is also simultaneously searched in multiple directions.

On SKA-Low, 500 tied-array beams will be formed, producing pulsar search data over a $\sim$100-MHz band sampled every $\sim$100 us,
while 1500 tied-array beams will be formed  on SKA1-Mid about 60 Petabytes/day data collected over a $\sim$300-MHz band sampled every $\sim$64 us. 

\begin{figure}[H]
    \centering
    \includegraphics[width=0.5\textwidth]{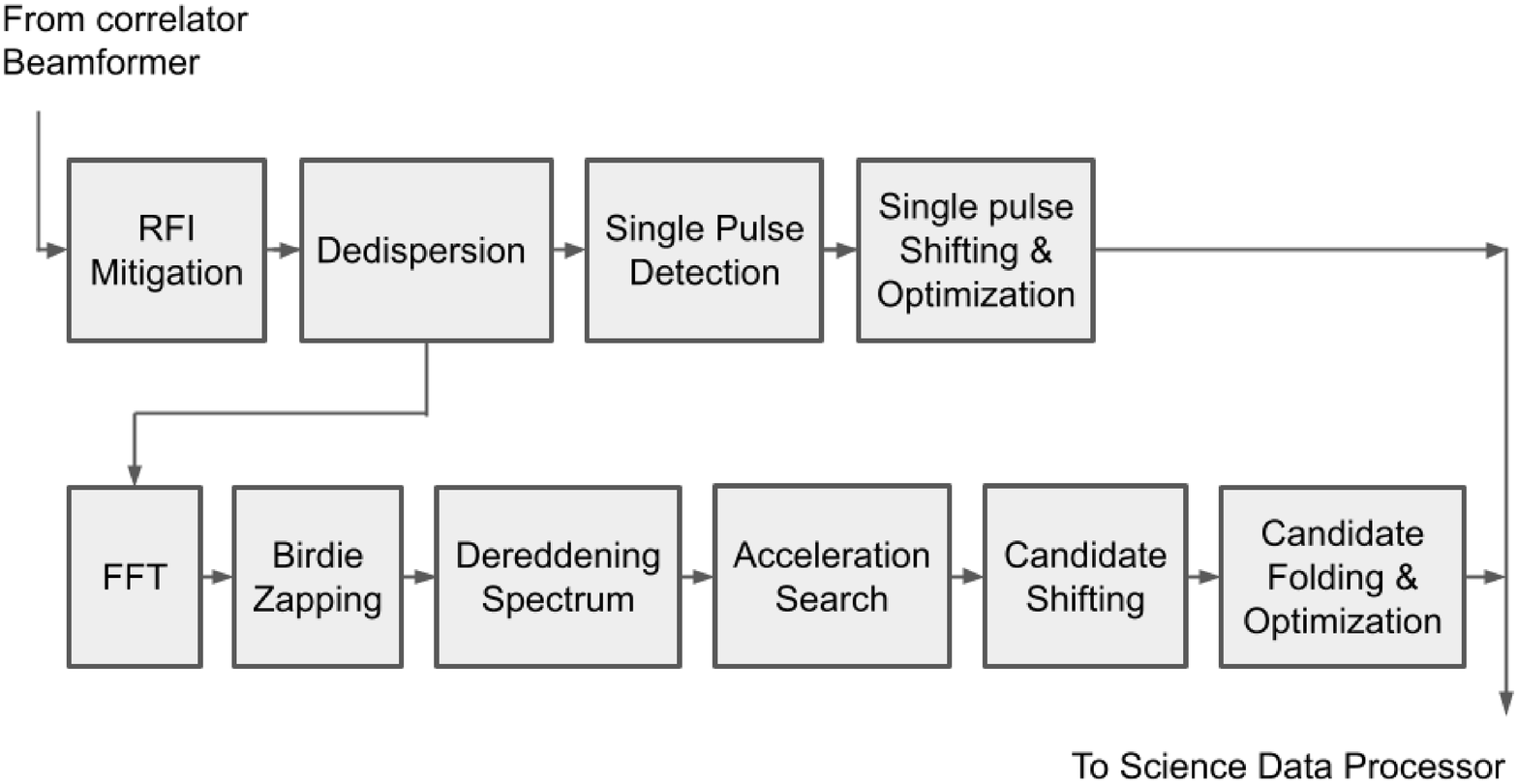}
        \caption{ Schematic of the pulsar search processing workflow in SKA. FPGA-based acceleration architecture investigated for 1) RFI mitigation, 2) dedispersion, 3) single pulse detection, 4) FFT, 5) acceleration search and 6) candidate folding $\&$ optimisation modules.}
        \label{fig:workflow}
\end{figure}

These requirements imply a need to solve a complex problem. The pulsar search involves 10 Peta-operations/sec. A detailed workflow of the SKA pulsar search pipeline is shown in Fig.~\ref{fig:workflow}. A  description of the various computing blocks shown in the figure is presented in \citep{levin}.


Based on the present technology, a computing solution for the pulsar search could need a huge compute cluster with 100s of nodes. Typically achieving higher numbers to the OPs/watt metric is desirable for high-performance computing. It will result in lower operational costs and lifetime for the compute elements. In the wake of the new technology era, FPGA-based acceleration could be easily evaluated. 

An FPGA accelerator on a host PC was used to assess their performance for pulsar search algorithms. The FPGA codes for the HPC are based on VHDL and OpenCL. To study the design development cycles for the FPGA-based accelerator platform, we have prototyped a few compute-intensive modules from the pulsar search application in the platform. FPGA design architectures, and in some cases, architectures and codes investigated for 

\begin{enumerate}
    \setlength{\itemsep}{2pt}  
    \setlength{\itemindent}{10pt}  
        \item Median filtering for RFI mitigation, 
        \item Dedispersion, 
        \item Single pulse detection, 
        \item Long FFT, 
        \item Acceleration search, and 
        \item Candidate folding optimisation modules
\end{enumerate}

Median filters with very large windows are investigated for use in the SKA pulsar search application. In this work, we have explored an approach suitable for operating on huge rectangular windows on an FPGA accelerator device and supporting the processing of infinite data streams. OpenCL-based codes were developed for this purpose and tested on Arria-10 FPGA accelerators \citep{sherwin}. 
A dedispersion module was developed in collaboration with the Altera (now Intel) group. OpenCL-based codes were created for this purpose and tested on Arria-10 FPGA accelerators. OpenCL-based programs were also designed for single-pulse detection and tested on Arria-10 FPGA accelerators. The long FFT modules were developed on Intel's off-chip example libraries. We could implement and verify 1, 4, 8 and 16 million-point FFTs on Arria-10 and Stratix-10 FPGA accelerators. Acceleration search is another complex module in the pulsar search pipeline. We have implemented and tested designs for the acceleration search both using VHDL and OpenCl on Arria-10 accelerators \citep{haomio1, haomio2, haomio3, haomio4, haomio5, haomio6, haomio7, fdas}. Finally, we have also made an architectural design study of the Candidate folding optimisation module for an Arria 10-based accelerator implementation.

Prototype investigations considered an observation period of about 10 s, time-resolution of 64 us, and a range of pulsar parameters. A list of SKA1 pulsar search parameters available in Table 1 of \citep{levin}. We have made progress in achieving our first results, especially, for the long FFT and in the acceleration search, we could achieve better than near real-time performance using the present technology FPGA accelerators accessible from three different vendors. We could also observe appreciable compute power per watt performance on these accelerators. 

\subsection{\textbf{SKA Low - Correlator and Beamformer}} \label{SkaLowCorrBeamformer}
\hfill

Correlators and beamformers are among the most complex signal processing modules in SKA \citep{bunton}. The SKA is considering different implementation choices for the mid and low-frequency telescopes. The SKA-mid telescope planned to use dedicated FPGA-based acceleration hardware for correlators and beamformers, while the SKA-low telescope is considering COTS-based FPGA accelerators \citep{grant}. In section~\ref{sec:radioTeleSignalProcessing}, Fig.~\ref{fig:twolayers}, we have seen that correlators and beamformers belong to the level-1 layer of signal processing, where dedicated hardware implementation usually is required. Due to the advancement of the COTS FPGA accelerators, this alternate implementation seems viable now. This section gives a brief review of the SKA Low correlator and beamformer architecture.

The SKA-Low telescope consists of 512 stations. Each station produces 384 frequency channels with a channel spacing of 781 kHz to provide a maximum bandwidth of 300 MHz. From 512 stations, up to 48 independent beams can be formed across 300 MHz bandwidth. In pulsar timing (PST) mode, with a frequency resolution of 14.5 kHz, 16 beams can be created with 300 MHz bandwidth. PST beams are used to time the known pulsars accurately. Pulsar search mode will have a frequency resolution of 3.6 kHz, and it is possible to form 500 beams with 118 MHz of bandwidth, and 250 beams with 236 MHz of bandwidth. The hardware investigated for the correlator beamforming consists of multiple servers with many PCIe FPGA accelerators. Most modern FPGAs with high-bandwidth (HBM) memories are explored as a potential accelerator element. 100 GbE ports on these accelerators would help with monitoring, hardware configuration and provide other local interfaces. Network switches are used for exchanging data. 
 
\section{\textbf{Ongoing and Future Opportunities}} \label{sec:OngoingAndFutureOpportunities}
\hfill

In this section, we review potential applications of FPGA-based acceleration with SKA pathfinders and other telescopes.

\subsection{\textbf{CASPER-based receivers}}
\label{sec:casper}
\hfill

The Collaboration for Astronomy Signal Processing and Electronics Research (CASPER) MSGE (Matlab, Simulink, System generator, embedded design kit tool flow) is an open-source tool for reconfiguring the FPGA data acquisition boards for radio astronomy. Mainly reconfigurable frontend ADC cards vary from single RF channel to 64 RF channel cards for a range of Virtex series and Kintex FPGAs. These FPGA boards can be programmed using MATLAB, Simulink, and Xilinx System Generator (XSG). All the FPGA boards have their controllers like PPC or Microblaze-based processors. FPGA boards were interconnected with a 10 Gb Ethernet interface for data transfer\citep{casper}.

Details of the CASPER-based digital receiver developed for the uGMRT are available in \citep{gupta, reddy}. A new receiver based on CASPER was developed for pulsar observation at the Gauribidanur observatory. The design details are discussed in \citep{banea}. 

\subsection{\textbf{OpenCL developments towards  eGMRT}}
\label{sec:fpgaAccelGMRT}
\hfill

The GMRT aided with a major upgrade of increasing the instantaneous bandwidth and providing almost seamless frequency coverage is now one of the most sensitive facilities (called uGMRT; \citep{gupta, reddy})
at low and mid frequencies in the pre-SKA era. In order to increase the collecting area of the GMRT at least by a factor of 2, a design study for the expansion of GMRT (called eGMRT; \citep{patra2019expanded}) was performed. In addition to providing an unprecedented sensitivity at low-frequency, the proposed plan of mounting 30-beam phased array feed (PAF) in each antenna, will result in an order of magnitude increase of the field-of-view, making eGMRT a promising survey instrument for pulsars and transients \citep{roy2017time}.  However, this warrants almost a two orders of magnitude increase in the compute cost (up to a few PetaFlops) compared to the uGMRT to build a correlator and beamformer for the eGMRT supporting few Tb/s of input data rate.  

Such systems need energy-efficient hardware with rapid development cycles. Rapid advances in VLSI technology over the past few decades have enabled FPGAs with the advantages of flexibility, low latency and power efficiency. Recent FPGAs provide higher Flops per watt making it an energy efficient hardware platform suitable for eGMRT. 
Since traditional HDLs involve longer development times, OpenCL was chosen as the development platform to enable faster development cycles.
A multi-station correlator is now being built on an Intel Arria-10 FPGA. Unlike GPUs, such a design also opens up the possibility of directly receiving the antenna signals on the accelerator using a on-board 40/100 GbE interface. Thus FPGA accelerators can also be used as a low-power system for pre-processing of the input data for applying radio frequency mitigation techniques, forming PAF beams etc.  

\subsection{\textbf{On the use of Random Forest ML tools to detect pulsar signatures}}
\label{sec:fpgaAccelRandForest}
\hfill

Traditionally standard pulsar analysis tools \citep{sigproc, dspsr, psrchive, presto}, and a few custom tools are mostly used for pulsar signal detection. While these tools have been successful in the detection of several transient objects, there are more challenges to be met when it comes to traditional tool usage with modern telescope data \citep{pulsarchallenges}. The idea of using Machine Learning (ML) techniques for transient candidate detection has been around for a long time, especially when handling larger and more varied sets of data. Some typical examples include algorithms such as Random Forest. The search for Pulsars and FRB using ML can comprise of two parts:   

Training from a known data source (containing both pulsars and non-pulsar signatures) and using this trained algorithm to test further unknown data.
The algorithm's training is heuristic-based learning, and the usage of machine learning algorithms (such as Random Forest) aid the process.

With the evolving technology, radio telescopes also have evolved. With the Square Kilometer Array (SKA) radio telescope,  the data volume is tremendous, and thus the need to process them faster has emerged. To combat this problem, we need a system that learns directly from the data, and this is where deep learning methods can be used effectively \citep{CNNforSKA, CNNImplementation}. 

With the increased data size and variety, CPU-based platforms have started using specialized accelerators to meet the ML tools processing requirement and handle associated computation challenges. Accelerators based on GPUs have been explored. The GPUs have high-performance libraries to execute numerical algorithms such as the FFTs run faster, and their programming model is suitable for exploiting the memory requirement.

Similarly, with significantly higher compute capability and a high amount of on-chip cache memory, the memory bottlenecks due to external memory access latency and power consumption are reduced in the case of the FPGA-based accelerators. In addition, the FPGAs support various data type precision such as the half-precision ( \href{https://www.intel.com/content/www/us/en/developer/articles/technical/intel-deep-learning-boost-new-instruction-bfloat16.html}{bfloat 16}\footnote{\url{https://www.intel.com/content/www/us/en/developer/articles/technical/intel-deep-learning-boost-new-instruction-bfloat16.html}}, besides single \& double-precision) floating-point number, which is significantly efficient to implement ML inference algorithms. Typically, FPGAs also show higher power efficiency on ML applications than their general-purpose computing counterparts \citep{FPGAadvantage}.

The amalgamation of FPGA-based accelerators and ML/deep learning would pave the way for future detection and the ability to handle large streams of live data: a challenge that is not yet completely solved \citep{FPGARF}.

Although various efforts have been taken in this direction, each implementation has its own merits, and demerits \citep{FPGACOMP}. The most common complexity is the resource and memory optimization for the ML suite training applications. Currently, most of the training is handled through existing software on general-purpose systems, and the testing/inference of new data through hardware accelerators \citep{FPGARF}. FPGA-based ML inference development frameworks are beginning to emerge \citep{Reuther}.

\subsection{\textbf{On the use of MaskRCNN ML tools to detect binary pulsar signatures}}
\label{sec:fpgaAccelMaskRcnn}
\hfill

Binary pulsar signatures appear with a distinct shape when the pulsar search pipeline's intermediate stage outputs are externally viewed. A study was made on using ML-based image recognition tools to detect these image patterns. A MaskRCNN ML suite was trained to detect these image patterns. A detailed account of this work is available in \citep{sbhat}. There are no publicly available Mask-RCNN inference tools for the direct implementation of modern FPGAs. Developing Mask-RCNN tools for the FPGA accelerators will enable handling of large amounts of data that SKA telescopes generate and enormously faster processing. We have found that primitive functions to build Mask R-CNN are available in OpenCL. However, with advanced building blocks in the OneAPI deep learning toolkit, Mask R-CNN  implementation and achieving accelerated inference technology on FPGA are expected to be faster. 

\subsection{\textbf{Emerging platforms}}
\label{sec:fpgaAccelEmerging}
\hfill

Modern FPGAs are revolutionizing the digital system designs with built-in analog to digital converters (RFSoC), ARM class processors, floating-point processing capabilities and high-level programming tool support. 

\subsubsection{\textbf{RFSoC FPGA-based accelerator}}
\label{sec:fpgaAccelRfsoc}
\hfill

A versatile PCIe accelerator board with high-speed digitizer (up to 32 analog signals) capabilities is being developed to meet the next generation tile processing requirements with correlation, beamforming for a multi-element radio telescope application \citep{girishetal}.


\subsubsection{\textbf{Integrated Accelerators}}
\hfill

With improvements in Multi-chip Module (MCM) and chiplet integration technologies, devices are able to host CPUs, FPGAs and other accelerators within the same package. Such heterogeneous devices could use strategies like Uniform Memory Access (UMA) which give the CPU and accelerators access to the same memory space, eliminating the need for transferring data between separate CPU and accelerator memories like in conventional systems. This would alleviate the host-accelerator data transfer bottleneck and improve energy efficiency. Similar strategies have been shown to be successful in newer hardware like in the Apple M1 SoC, which provides the CPU, integrated GPUs and other accelerators access to the same memory pool. Similar devices with FPGAs have also been introduced in the past with Intel releasing a Xeon Skylake Processor with an integrated Arria 10 FPGA, although it should be noted that this device has a Non-UMA memory architecture.


A recent work \citep{mv2021riska} also examines the merits of such approaches for SKA and proposes a custom SoC integrated with domain-specific architectures for such demanding algorithms to meet its performance and energy efficiency requirements.



\subsubsection{\textbf{Open source landscape}}
\hfill

Most FPGA tool-chains presently are available as licensed proprietary software supplied by the FPGA vendors, leading to additional development costs and separate ecosystems with little interoperability. Initiatives like the FOSS Flow For FPGA (F4PGA) project are trying to change this. It is, among other things, developing free and open-source toolchains for programming FPGAs and introducing the FPGA interchange format to improve interoperability between toolchains. While the rate of adoption of such toolchains will depend on their pace of reaching maturity and support availability for different FPGA platforms, such initiatives will enrich and improve the FPGA tooling ecosystem. Apart from the FPGA tooling, open source IPs have also been gaining in popularity. With open ISAs like RISC-V and open interconnect standards, many verified component libraries including general-purpose cores, arithmetic units, and interconnects are being developed as open source projects. Leveraging such libraries will reduce development times as well as allow more robust implementations.


 
\section{Summary}  


This paper presented an overview of the need for high-performance computing for radio-telescope signal processing problems through background information and case studies. We have introduced the signal processing problems of a modern telescope in two layers (Fig.~\ref{fig:twolayers}). Very high-speed and purpose-built hardware platforms traditionally handle Layer-1 of the problem. Layer-2 signal processing, such as the pulsar search, correlation-beamforming etc., were routinely done offline. 

We have also presented in section \ref{sec:challRadioTeleSignalProcessing}, a glimpse of the significant challenges of radio astronomy signal processing and discussed the need for high memory throughput, high input-output throughput and high performance per power unit. Subsequently, we have presented an illustration to show that achieving high performance per power-unit is a complex problem.  
 
In section \ref{sec:accelLand}, we have provided a historical overview of the accelerators built for high-performance computing. FPGAs are one of the new accelerators that entered the COTS accelerator market. FPGAs have large arrays of parallel logic, memory and signal processing modules. Hence, they offer purpose-built hardware to implement and execute complex algorithms more efficiently. This efficiency is possible due to the fine-grain parallel nature of the FPGA fabric and its reconfigurability  (Fig.  \ref{fig:fpga}).

We have provided a historical overview of the FPGA PCI(e) accelerator and show its architectural similarity with the present-day COTS FPGA accelerators in section \ref{sec:fpgaAccel}.

We show that the FPGAs are moving into the high-performance computing accelerator market (Fig. \ref{fig:plx}). We also observe that the firmware architecture layers and the code development tools of the COTS accelerators are evolving to provide a seamless interface for the code development and house them in multiple host environments. We have a top-level review of the evolving FPGA-accelerator's firmware architecture, presenting details of how reconfigurability is achieved in sections \ref{sec:fpgaAccelArch} \& \ref{sec:fpgaAccelReconfig}. 
 FPGA accelerator code development procedures are different from the conventional accelerator tools such as the GPUs, and they are reviewed in sections \ref{sec:fpgaAccelCodeDev}, \ref{sec:HighLevelLanguageTools} \& \ref{sec:DevelopmentEnvironment}.  In these sections, we have discussed the pros and cons of both HDL and high-level tools and present a balanced view. 


The SKA project focuses on using suitable and efficient technology for ultra-fast computing and handling and distributing the enormous amounts of data the telescope generates. It implements a complex, traditionally offline executed layer-2 type signal processing for pulsar search algorithms to run in real-time (in section \ref{pulsarserach}). In addition, we have also shown an example case from the SKA Low, where the traditional layer-1 problem is being investigated for real-time implementation using FPGA accelerators (in section \ref{SkaLowCorrBeamformer}).
\hfill

We have also presented in section \ref{sec:OngoingAndFutureOpportunities}, details of a few other ongoing activities where FPGA-based accelerators are already in use. One of the works is based on CASPER tools for implementing a layer-1 signal processing at the Gauribidanur radio telescope facility (section \ref{sec:casper}). A second work is about the implementations for the uGMRT, where CASPER FPGA tools and hardware are used. Another work is the ongoing OpenCL-based efforts to implement the layer-1 computations for the eGMRT work involving a design for phased array feed (section \ref{sec:fpgaAccelGMRT}. 

Subsequently, in sections \ref{sec:fpgaAccelRandForest} \& \ref{sec:fpgaAccelMaskRcnn}, we have presented details of the machine learning problems, especially for the compute problems related to pulsar candidate detection, where traditionally direct human interactions were needed. The focus of this section is to investigate FPGA-based solutions to implement inferencing algorithms that need to deal with high-volume data streams, such as what will come from the square kilometre telescopes.


 Our review of emerging platforms in section \ref{sec:fpgaAccelEmerging} highlights the upcoming modern RFSoC-based FPGA accelerator in-house developments presented in this journal. Further, we reviewed industry efforts in integrating the FPGA \& CPU and their approach to providing unified access to the main memory for the accelerator. FPGA accelerator performance depends on both hardware and firmware. Lastly, we reviewed the landscape of the open-source FPGA tools and their potential use in developing purpose-built hardware in the future.

\section{Conclusion}
\hfill

This paper reviewed FPGA-based acceleration of the SKA non-imaging pipeline, particularly for the pulsar search, and the details of the pulsar timing pipeline are discussed elsewhere in \citep{barr}. We have reviewed the SKA Low correlator beamformer FPGA-based acceleration plans. The SKA Mid correlator beamformer is also  built using dedicated FPGA-based hardware and the details are available in \citep{Gerhard}. SKA science data processing computers will receive about 5 Terabits of data per second. Relevant details of the SKA imaging pipeline-related compute profile complexity investigations are listed elsewhere in \citep{vermijChallenges2015, jongeriusExascale2016}. The SKA science regional centres will also host several high-performance computers, and details about their complexity and early investigations are available elsewhere in \citep{Garrido}

In general, the FPGA-based accelerators appear attractive due to their high power performance \citep{Parker2021, Qasaimeh} \& as seen in Fig.\ref{fig:gfloppw}, leading to environment-friendly green-computing\footnote{\url{https://en.wikipedia.org/wiki/Green_computing}} options. 

In addition, advancement from the FPGA vendors towards providing: a)  faster design synthesis tools/techniques, b) partial/dynamic reconfigurability, c) incremental compilation, d) libraries to mix and match at the FPGA binary levels, e) availability of just-in-time compilers, f) archives to help code reuse and g) archives/web-servers for extended lifetime support of the development environments would further enhance the acceleration performance for a larger class of high-performance computing problems with FPGA-based accelerators.

FPGA based accelerators are emerging technology in the high performance landscape. Complex algorithms benefit from the fine grain parallelism and power efficiencies offered by the FPGA accelerators. We anticipate benefits using FPGA-based acceleration in the \textit{Square Kilometre Array} project \textit{as well in the upcoming high performance applications}. 



\section*{Acknowledgements}
This paper presents a culmination of ideas over years of work and exchanging ideas through various formal and informal discussions and meetings. We acknowledge those interactions and all support received towards it, especially from several other colleagues and students from our workplaces. We also thank the FPGA industry colleagues to whom we could talk to get a broader picture of the acceleration and high-performance scenarios from their perspective. And, finally we thank the anonymous reviewers whose comments helped to shape this paper.  
\pagebreak

\setlength{\tabcolsep}{20pt}
\setlength{\columnsep}{2cm}
\renewcommand{\arraystretch}{1.5}


\vspace{-1.5em}
\newcommand{\comment}[1]{}
\comment{

}

\bibliography{bibliography}
\end{document}